\documentclass[aip,amsmath,amssymb,reprint]{revtex4-1}
\usepackage{graphicx}
\usepackage{dcolumn}
\usepackage{bm}

\begin{document}

\title{Configurational entropy of polydisperse supercooled liquids}

\author{Misaki Ozawa}

\affiliation{Laboratoire Charles Coulomb (L2C),
University of Montpellier, CNRS, Montpellier, France}

\author{Giorgio Parisi}

\affiliation{Dipartimento di Fisica, Universit\`a degli studi di Roma La Sapienza, Nanotec-CNR, UOS Rome, INFN-Sezione di Roma 1, Piazzale A. Moro 2, 00185, Rome, Italy}

\author{Ludovic Berthier}

\affiliation{Laboratoire Charles Coulomb (L2C),
University of Montpellier, CNRS, Montpellier, France}

\begin{abstract}
We propose a computational method to measure the configurational entropy in generic polydisperse glass-formers. In particular, our method resolves issues related to the diverging mixing entropy term due to a continuous polydispersity. The configurational entropy is measured as the difference between the well-defined fluid entropy and a more problematic glass entropy. We show that the glass entropy can be computed by a simple generalisation of the Frenkel-Ladd thermodynamic integration method, which takes into account permutations of the particle diameters. This approach automatically provides a physically meaningful mixing entropy for the glass entropy, and includes contributions that are not purely vibrational. The proposed configurational entropy is thus devoid of conceptual and technical difficulties due to continuous polydispersity, while being conceptually closer, but technically simpler, than alternative free energy approaches.
\end{abstract}

\maketitle

\section{Introduction}

\label{sec:Intro}

Polydispersity is an essential ingredient to study supercooled liquids and glasses because mono-component glass-forming systems with spherical particles quickly crystallize and do not easily form amorphous states.
For example, it is well-known that multi-components metallic glasses with sufficiently large size polydispersity show better glass-forming ability~\cite{chen2011brief}, and size polydispersity is unavoidable in colloidal glasses~\cite{hunter2012physics}. Continuously polydisperse glass-forming models are also getting increasing attention because they maximise the efficiency of the swap Monte-Carlo algorithm~\cite{gazzillo1989equation,frenkel2001understanding,grigera2001fast}. As a result, they can be equilibrated down to extremely low-temperatures or large densities~\cite{ninarello2017models,berthier2016equilibrium}.
This recent computational development enables numerical studies that can be directly compared to experimental work, and opens several possibilities to explore a wide range of physical phenomena occurring in amorphous materials~\cite{berthier2017configurational,berthier2016growing,ozawa2018random}. 

A central issue for supercooled liquids is the determination of their configurational entropy, and of its evolution when approaching the glass transition\cite{berthier2011theoretical}. However, the statistical mechanics of continuously polydisperse systems involves some controversial issues such as particle distinguishability and the associated divergent mixing entropy~\cite{warren1998combinatorial,swendsen2006statistical,maynar2011entropy,frenkel2014colloidal,paillusson2014role,cates2015celebrating,paillusson2018gibbs}.
These issues also influence the statistical mechanics description of polydisperse glass-formers~\cite{ozawa2017does}. The configurational entropy $S_{\rm conf}$ can be defined by the difference between the total entropy, $S_{\rm tot}$, and a glass entropy, $S_{\rm glass}$, 
\begin{equation}
S_{\rm conf} = S_{\rm tot} - S_{\rm glass},
\label{eq:def}
\end{equation}
so that $S_{\rm conf}$ enumerates the number of glass states. The technical problem with Eq.~(\ref{eq:def}) is evident as we need to take the difference between two entropies evaluated separately in phases that are not connected by any equilibrium thermodynamic path. The unwanted byproduct is that the {\it absolute values} of both entropies are needed. This is particularly problematic for continuously polydisperse models, since the entropy $S_{\rm tot}$ then contains a mixing entropy contribution that is formally divergent, while conventional methods to determine $S_{\rm glass}$ do not. As a result, widely-used methods to determine $S_{\rm conf}$ in systems with continuous polydispersity provide an infinite value, which is unphysical. Similar problems are encountered by discrete mixtures with infinitesimal size differences, where the mixing entropy contribution to glass and fluid entropies is again a problematic issue~\cite{ozawa2017does}. It is therefore important to develop methods to properly deal with the mixing entropy contribution to $S_{\rm glass}$ in Eq.~(\ref{eq:def}), so that meaningful configurational entropy measurements can be generically performed for any type of particle size distributions with no ad-hoc manipulations of mixing entropy contributions. The main goal of the present paper is to provide such a computational method.

For ordinary phase transitions, only {\it entropy differences} are physically relevant, and can be measured by following an equilibrium thermodynamic path between two state points. This is how experiments get around the absolute value problem for glasses, too, but as a result only an approximate estimate of the configurational entropy can be measured~\cite{richert1998dynamics,angell2002specific,tatsumi2012thermodynamic}. 
In a previous article~\cite{ozawa2017does}, we provided a resolution to the problem of the infinite mixing entropy contribution to Eq.~(\ref{eq:def}). 
The key physical idea is that glass configurations that only differ by the exchange of particles with very similar sizes should be considered as part of the same glass `state' and must be grouped together when estimating $S_{\rm glass}$. This suggests that a glass state is associated with an infinitely large number of configurations, and thus $S_{\rm glass}$ contains a divergent mixing entropy contribution term which cancels the one in $S_{\rm tot}$, to eventually make $S_{\rm conf}$ finite. In Ref.~\onlinecite{ozawa2017does}, we provided an approximate method  to evaluate a finite $S_{\rm conf}$, which amounts to describing a continuously polydisperse system as an effective discrete mixture with a finite number of species, $M^*$. We proposed an empirical method to estimate $M^*$ directly in the simulations for each state point, and applied this approach to a number of glass-formers~\cite{ozawa2017does,berthier2017configurational}.    
However, a general and precise treatment of the mixing entropy is desired that does not rely on approximations and can also be applied to an arbitrary functional form of the particle size distribution. This is becoming a particularly pressing issue as computer simulations are now getting closer to a putative thermodynamic transition, which is defined by a vanishing configurational entropy. Thus, it is no longer possible to work with empirical, approximate methods to address the nature of the glass transition.
As argued in our previous paper~\cite{ozawa2017does}, the mixing entropy of the glass state needs to be included in Eq.~(\ref{eq:def}), since failure to do so leads to the incorrect conclusion~\cite{baranaunew} that the configurational entropy is bounded from below by the mixing entropy. 

The goal of this paper is to provide a proper statistical mechanics description and a generic computational scheme to obtain the configurational entropy of continuously polydisperse systems. We thus transform the empirical method and the physical ideas proposed in Ref.~\onlinecite{ozawa2017does} into a mathematically consistent computational scheme applicable to any type of particle size distribution. 
The computational method that we establish in this work relies again on Eq.~(\ref{eq:def}), but we use a statistical mechanics description of $S_{\rm glass}$ that includes particle permutation, and thus automatically produces the correct mixing entropy. Whereas the evaluation of $S_{\rm tot}$ remains unchanged, 
$S_{\rm glass}$ is now computed by a Frenkel-Ladd thermodynamic integration~\cite{frenkel1984new} that we generalize to deal with the mixing entropy. To demonstrate that our method provides physically meaningful results, we perform molecular dynamics simulations of three glass-forming models, using continuously polydisperse soft and hard spheres~\cite{ninarello2017models,berthier2016equilibrium},
and a binary Lennard-Jones mixture~\cite{kob1995testing}. Remarkably the obtained $S_{\rm conf}$ for the polydisperse hard spheres takes values comparable to the Landau free energy approach~\cite{berthier2014novel} based on the Franz-Parisi potential~\cite{franz1997phase}. This suggests that our scheme provides a cheaper computational alternative to free energy measurements.

This paper is organised as follows.
In Sec.~\ref{sec:framework}, we describe the general framework leading to our computational method.
Its numerical implementation for three representative glass-formers is presented in Sec.~\ref{sec:numerical}.
Finally, we conclude and discuss our work in Sec~\ref{sec:conclusion}.

\section{Statistical mechanics framework}

\label{sec:framework}

\subsection{Setting}

We consider an $M$-component polydisperse system in the canonical ensemble in $d$-dimensions, such that $N$, $V$, and $T=1/\beta$ are the number of particles, volume, and temperature, respectively. We fix the Boltzmann constant to unity, and $\rho=N/V$ is the number density. The case $M = N$ corresponds to a continuously polydisperse system. The concentration of the $m$-th species is $X_m=N_m/N$, where $N_m$ is the number of particles of the $m$-th species ($N=\sum_{m=1}^M N_m$). A point in position space is denoted as ${\bf r}^N=({\bf r}_1, {\bf r}_2, \cdots, {\bf r}_N)$. For simplicity, we consider equal masses, irrespective of the species.

\subsubsection{Partition functions}

For $M$-component polydisperse systems, the following partition function in the canonical ensemble is conventionally used~\cite{frenkel2014colloidal}:
\begin{equation}
Z = \frac{1}{\Pi_{m=1}^M N_m! \Lambda^{Nd}} \int_V \mathrm{d} {\bf r}^N e^{-\beta U ({\bf r}^N)},
\label{eq:partition_function_old}
\end{equation}
where $\Lambda=\sqrt{2\pi \beta \hbar^2/m}$ and $U ({\bf r}^N)$ are the de Broglie thermal wavelength and the potential energy, respectively. We set the mass $m=1$ and the Planck constant $\hbar=1$. Note that in Eq.~(\ref{eq:partition_function_old}), the position ${\bf r}^N$ is the only pertinent degree of freedom left after tracing out the momentum.

For polydisperse systems, it is however useful to consider the permutation of the particle diameters as additional degrees of freedom.
We define a set of diameter $\Sigma^N$ as $\Sigma^N=\{ \sigma_1, \sigma_2, \cdots, \sigma_N  \}$.
We introduce a permutation $\pi$ to the set $\Sigma^N$, and
$\Sigma_{\pi}^N$ represents a specific sequence of the diameters, e.g., $\Sigma_{\pi}^N=(\sigma_3, \sigma_8, \sigma_5,\cdots)$. 
In total there exists $N!$ such permutations.
We define a reference sequence, $\Sigma_{\pi^*}^N=(\sigma_1, \sigma_2, \sigma_3, \cdots, \sigma_N)$.
Now the potential energy also depends on the permutation $\pi$ as denoted by $U (\Sigma_{\pi}^N, {\bf r}^N)$. For simplicity, we write $U ({\bf r}^N)=U (\Sigma_{\pi^*}^N, {\bf r}^N)$ only for the reference $\Sigma_{\pi^*}^N$ and drop off $\Sigma_{\pi^*}^N$ from the argument.

Because we include the permutations as additional degrees of freedom, we sum up all the possible permutations in the partition function as 
\begin{equation}
\mathcal{Z} =  \frac{1}{N!} \sum_{\pi} \frac{1}{\Pi_{m=1}^M N_m! \Lambda^{Nd}} \int_V \mathrm{d} {\bf r}^N  e^{-\beta U (\Sigma_{\pi}^N, {\bf r}^N)}.
\label{eq:partition_function_new}
\end{equation}
This generalised partition function in Eq.~(\ref{eq:partition_function_new}) is the correct starting point to compute the total and glass entropies.

\subsubsection{Frenkel-Ladd Hamiltonian}

We denote the potential energy of the target system by $\beta U_0(\Sigma_{\pi}^N, {\bf r}^N)$.
To evaluate the entropy of the glass state by a Frenkel-Ladd thermodynamic integration~\cite{frenkel1984new,coluzzi1999thermodynamics,sastry2000evaluation,angelani2007configurational}, we need to impose a harmonic constraint with the spring constant $\alpha$ on the target system $\beta U_0(\Sigma_{\pi}^N, {\bf r}^N)$ as described by
\begin{equation}
\beta U_{\alpha}(\Sigma_{\pi}^N, {\bf r}^N, {\bf r}_0^N) = \beta U_0(\Sigma_{\pi}^N, {\bf r}^N) + \alpha \sum_{i=1}^{N} | {\bf r}_i - {\bf r}_{0 i} |^2,
\label{eq:hamiltonian}
\end{equation}
where ${\bf r}_0^N$ is a reference equilibrium configuration drawn from the Boltzmann distribution of the target system. We will use $\beta U_0(\Sigma_{\pi}^N, {\bf r}^N)$ and $\beta U_{\alpha}(\Sigma_{\pi}^N, {\bf r}^N, {\bf r}_0^N)$ (with $\alpha>0$) to access the total entropy and the glass entropy, respectively.

Note that in this approach, ${\bf r}_0^N$ is a randomly chosen equilibrium configuration of the fluid~\cite{angelani2007configurational,foffi2005,coluzzi1999thermodynamics}, so that the Frenkel-Ladd method implicitly assumes that the vibrational entropy associated with any reference configuration belonging to a given metabasin is the same for all configurations of that metabasin, and inherent structures play no specific role in that scheme. 

\subsection{Computing the total entropy $S_{\rm tot}$}

In this section we explain how to compute the total entropy $S_{\rm tot}$, starting from the partition function in Eq.~(\ref{eq:partition_function_new}). 

\subsubsection{A trivial identity}

The partition function in Eq.~(\ref{eq:partition_function_new}) of the target system defined by $\beta U_0(\Sigma_{\pi}^N, {\bf r}^N)$ reduces to the conventional partition function in Eq.~(\ref{eq:partition_function_old}) because permutations of diameters are always compensated by permutations of the positions if there is no constraint, namely
\begin{eqnarray}
\mathcal{Z}_0 &=&  \frac{1}{N!} \sum_{\pi} \frac{1}{\Pi_{m=1}^M N_m! \Lambda^{Nd}} \int_V \mathrm{d} {\bf r}^N  e^{-\beta U_0 (\Sigma_{\pi}^N, {\bf r}^N)} \nonumber \\
&=& \frac{1}{\Pi_{m=1}^M N_m! \Lambda^{Nd}} \int_V \mathrm{d} {\bf r}^N e^{-\beta U_0 ({\bf r}^N)} = Z_0.
\label{eq:partition_function_total}
\end{eqnarray}
Therefore, the computation of $S_{\rm tot}$ is not altered by the newly introduced summation associated with the permutations in Eq.~(\ref{eq:partition_function_new}). 

\subsubsection{Thermodynamic integration from the ideal gas}

Following the convention~\cite{sciortino1999inherent,sastry2000evaluation,coluzzi2000lennard,angelani2007configurational}, we perform a thermodynamic integration from the ideal gas state to the target state.
The thermodynamic integration for $S_{\rm tot}$ depends on the type of interaction potentials, and we need to distinguish between continuous potentials (`Soft') and hard sphere potentials (`Hard'). The resulting expressions are:
\begin{eqnarray}
S_{\rm tot} &=& S_{\rm id} + \beta E_{\rm pot}(\beta) - \int_0^{\beta} \mathrm{d} \beta' E_{\rm pot}(\beta') \quad {\rm (Soft)}, \label{eq:total_soft} \\
S_{\rm tot} &=& S_{\rm id} - N \int_0^{\phi} \mathrm{d} \phi' \frac{(p(\phi')-1)}{\phi'} \quad {\rm (Hard)},
\label{eq:total_hard}
\end{eqnarray}
where $S_{\rm id}$, $E_{\rm pot}$, $\phi$ and $p$ are the ideal gas entropy, the averaged potential energy, the volume fraction, and the reduced pressure, respectively. For the ideal gas, 
$S_{\rm id}$ can be written as
\begin{equation}
S_{\rm id}= N \frac{(d+2)}{2} - N \ln \rho - N \ln \Lambda^d + S_{\rm mix}^{(M)}, 
\label{eq:entropy_ideal_gas}
\end{equation} 
where $S_{\rm mix}^{(M)}$ is the mixing entropy of the ideal gas expressed as 
\begin{equation}
S_{\rm mix}^{(M)} = \ln \left( \frac{N!}{\Pi_{m=1}^M N_m!}\right).
\label{eq:s_mix_M}
\end{equation}
When $M$ is finite and $N_m \gg 1$, we can apply Stirling's approximation, $\ln N_m! \simeq N_m \ln N_m -N_m$, and  then Eq.~(\ref{eq:s_mix_M}) reduces to the standard form of the mixing entropy, $S_{\rm mix}^{(M)}/N=-\sum_{m=1}^M X_m \ln X_m$.

One can see that in a continuous polydisperse limit (where $M=N$, and hence $N_m=1$), $S_{\rm mix}^{(M)}$ diverges in the thermodynamic limit~\cite{salacuse1982polydisperse,frenkel2014colloidal}, $S_{\rm mix}^{(M=N)}/N=(\ln N!)/N \simeq \ln N -1  \to \infty$.
This divergence is the root of a paradoxical situation in the context of the glass physics as the divergence of $S_{\rm mix}^{(M)}$ would cause the divergence of $S_{\rm tot}$ and hence $S_{\rm conf}$, suggesting that the glass transition may not happen~\cite{ozawa2017does,baranau2017another}.

\subsection{Computing the glass entropy $S_{\rm glass}$}

We compute the entropy of the glass state, $S_{\rm glass}$, by a Frenkel-Ladd construction~\cite{frenkel1984new,coluzzi1999thermodynamics,sastry2000evaluation,angelani2007configurational}, starting from Eq.~(\ref{eq:partition_function_new}) with $\beta U_{\alpha}(\Sigma_{\pi}^N, {\bf r}^N, {\bf r}_0^N)$ ($\alpha>0$) in Eq.~(\ref{eq:hamiltonian}).
The central idea of the Frenkel-Ladd construction is to perform a thermodynamic integration between a well-known limit, the Einstein solid when $\alpha$ is very large and particles perform small vibrations around the positions dictated by the reference configuration to small $\alpha$ where the vibrations resemble the ones of the glass. This thermodynamic path involves an integration of the mean squared displacement from large to small $\alpha$-values. We now explain this process.

\subsubsection{Partition function in glass state}

For the glass state $\alpha$ defined by the vicinity of the reference configuration, the partition function in Eq.~(\ref{eq:partition_function_new}) becomes
\begin{equation}
\mathcal{Z}_{\alpha} =  \frac{1}{N!} \sum_{\pi} \frac{N!}{\Pi_{m=1}^M N_m! \Lambda^{Nd}} \int_V \mathrm{d} {\bf r}^N e^{-\beta U_{\alpha}(\Sigma_{\pi}^N, {\bf r}^N, {\bf r}_0^N)}.  \label{eq:Z_alpha}
\end{equation}
We add a factor $N!$ in the numerator of Eq.~(\ref{eq:Z_alpha}), because for a given reference configuration ${\bf r}_0^N$, there exist $N!$ exactly identical configurations defined by the corresponding permutations of the particle identities, which we must take into account (see Ref.~\onlinecite{coluzzi1999thermodynamics} for a related argument). 
Note that due to the presence of the reference configuration ${\bf r}_0^N$, the identity shown in Eq.~(\ref{eq:partition_function_total}) does not hold in the glass state.

We can then compute the entropy $S_{\alpha}$ by $S_{\alpha}=\beta E_{\alpha} - \beta F_{\alpha}$, where $E_{\alpha}$ and $F_{\alpha}=- \beta^{-1} \ln \mathcal{Z}_{\alpha}$ are the total energy and free energy of the state $\alpha$, respectively.

\subsubsection{Definition of glass entropy}

We define the glass entropy of the target system as follows:
\begin{equation}
S_{\rm glass} = \lim_{\alpha_{\rm min} \to 0} \overline{S_{\alpha_{\rm min}}},
\label{eq:S_glass_def}
\end{equation}
where $\overline{(\cdots)}$ represents a (disorder) average over the reference configuration ${\bf r}_0^N$ defined in Eq.~(\ref{eq:template_average}) below.

The limit operation, $\lim_{\alpha_{\rm min} \to 0}$, is crucial both conceptually and practically. Although the naive limit leads back to the fluid state, here we wish to compute the entropy of a metastable glassy state characterised by a finite lifetime. To this end, we need to keep $\alpha_{\rm min}$ finite, to prevent the exploration of a different glass state during the thermodynamic integration, and we instead make a simple extrapolation of $\alpha_{\rm min}$ from a finite $\alpha_{\rm min}$ value where a metastable glass state is well-defined, down to zero. This kind of extrapolation is inevitable in handling metastable states in finite dimensions, which all have a finite lifetime.
Our practical solution to accurately perform the limit is explained below in Sec.~\ref{sec:numerical}. 

We pick up the reference configuration ${\bf r}_0^N$ from equilibrium configurations drawn from the Bolzmann distribution of the target system.
This choice makes our scheme conceptually closer to the Franz-Parisi free energy approach in that the overlap function is computed using equilibrium reference configurations~\cite{franz1997phase,berthier2014novel}.
One might intuitively think that configurations at the inherent structure would be natural candidates for ${\bf r}_0^N$.
However, the present choice produces quantitatively consistent results with a vibrational description around inherent structures as confirmed in the Kob-Andresen model~\cite{ozawa2018ideal} and polydisperse soft spheres (Fig.~\ref{fig:FL}(b)).
Thus, we expect that equilibrium reference configurations ${\bf r}_0^N$ inside a basin of attraction produce essentially the same result as its inherent structure.

\subsubsection{Statistical averages} 

For convenience, we define the following notations of the various statistical averages needed in the different computations:
\begin{eqnarray}
\left\langle (\cdots) \right\rangle_{\alpha}^{\rm T, S} &=& \frac{\frac{1}{N!} \sum_{\pi} \int_V \mathrm{d} {\bf r}^N  (\cdots) e^{ -\beta U_{\alpha}(\Sigma_{\pi}^N, {\bf r}^N, {\bf r}_0^N)}  }{\frac{1}{N!} \sum_{\pi} \int_V \mathrm{d} {\bf r}^N  e^{-\beta U_{\alpha}(\Sigma_{\pi}^N, {\bf r}^N, {\bf r}_0^N)}}, \label{eq:T_S} \\
\left\langle (\cdots) \right\rangle_{\alpha}^{\rm T} &=& \frac{\int_V \mathrm{d} {\bf r}^N  (\cdots) e^{ - \beta U_{\alpha}({\bf r}^N, {\bf r}_0^N)}   }{\int_V \mathrm{d} {\bf r}^N  e^{ - \beta U_{\alpha}({\bf r}^N, {\bf r}_0^N)}  }, \label{eq:T} \\
\left\langle (\cdots) \right\rangle_{\beta}^{\rm S} &=& \frac{ \frac{1}{N!} \sum_{\pi} (\cdots) e^{ -\beta U_0(\Sigma_{\pi}^N, {\bf r}_0^N) }  }{ \frac{1}{N!} \sum_{\pi}  e^{ -\beta U_0(\Sigma_{\pi}^N, {\bf r}_0^N) }}, \label{eq:S} \\
\overline{(\cdots)} &=& \frac{\int_V \mathrm{d} {\bf r}_0^N  (\cdots) e^{ -\beta U_0({\bf r}_0^N) }}{\int_V \mathrm{d} {\bf r}_0^N  e^{ -\beta U_0({\bf r}_0^N) }}, \label{eq:template_average}
\end{eqnarray}
where the superscripts, T and S, represent the statistical average over positions (T) and permutations (S), respectively.
Numerically, these statistical averages can be easily evaluated through Monte-Carlo simulations using standard translational displacement (T) and particle swaps (S)~\cite{kranendonk1991free}.
Note that any permutation $\pi$ of the particle diameters can be expressed as a product of two-particle diameter swaps, and thus the permutation-phase space can be properly sampled using swap Monte-Carlo simulations.

\subsubsection{Large $\alpha$-regime: Einstein solid}

In the Frenkel-Ladd construction, the Einstein solid is chosen as the reference state~\cite{frenkel1984new}.
When $\alpha_{\rm max}$ is very large, the system is constrained near the reference configuration ${\bf r}_0^N$, thus we get
$\beta U_{\alpha_{\rm max}}(\Sigma_{\pi}^N, {\bf r}^N, {\bf r}_0^N) \simeq \beta U_0(\Sigma_{\pi}^N, {\bf r}_0^N) + \alpha_{\rm max} \sum_{i=1}^{N} | {\bf r}_i - {\bf r}_{0 i} |^2$.
Therefore, using Eq.~(\ref{eq:Z_alpha}), the system is described by the Einstein solid whose free energy is given by
\begin{equation}
\beta F_{\alpha_{\rm max}} =  N \ln \Lambda^d  + W({\bf r}_0^N, \beta) + \frac{N d}{2} \ln \left(\frac{\alpha_{\rm max}}{\pi}\right) -S_{\rm mix}^{(M)},
\label{eq:F_alpha_max}
\end{equation}
where $W({\bf r}_0^N, \beta)$ is an effective potential defined by
\begin{equation}
W({\bf r}_0^N, \beta) = - \ln \left( \frac{1}{N!} \sum_{\pi} e^{-\beta U_0(\Sigma_{\pi}^N, {\bf r}_0^N)} \right).
\end{equation}
This term, which originates from the effect of the permutation, plays an important role in the evaluation of the mixing entropy of the glass state. This is discussed further below.

\subsubsection{Small $\alpha$-regime} 

We compute $S_{\alpha_{\rm min}}$ in Eq.~(\ref{eq:S_glass_def}) by $S_{\alpha_{\rm min}} =  \beta E_{\alpha_{\rm min}} - \beta F_{\alpha_{\rm min}}$, where $\beta E_{\alpha_{\rm min}}$ and $\beta F_{\alpha_{\rm min}}$ are respectively given by 
$\beta E_{\alpha_{\rm min}} =  \frac{N d}{2} + \beta \left\langle U_{\alpha_{\rm min}} (\Sigma_{\pi}^N, {\bf r}^N, {\bf r}_0^N) \right\rangle_{\alpha_{\rm min}}^{\rm T,S}
$
and a thermodynamic integration of the mean-squared displacement over $\alpha$,
\begin{equation}
\beta F_{\alpha_{\rm min}} = \beta F_{\alpha_{\rm max}} - \int_{\alpha_{\rm min}}^{\alpha_{\rm max}} \mathrm{d} \alpha \left\langle \sum_{i=1}^{N} | {\bf r}_i - {\bf r}_{0 i} |^2 \right\rangle_{\alpha}^{\rm T,S}.
\label{eq:F_TI}
\end{equation}
Therefore, together with Eq.~(\ref{eq:F_alpha_max}), we can express $S_{\alpha_{\rm min}}$ as
\begin{eqnarray}
S_{\alpha_{\rm min}}  &=&  \frac{Nd}{2} - N \ln \Lambda^d - \frac{Nd}{2} \ln \left(\frac{\alpha_{\rm max}}{\pi}\right) + \ S_{\rm mix}^{(M)} \nonumber \\
&\quad& -  W({\bf r}_0^N, \beta) + \beta \left\langle U_{\alpha_{\rm min}} (\Sigma_{\pi}^N, {\bf r}^N, {\bf r}_0^N) \right\rangle_{\alpha_{\rm min}}^{\rm T,S} \nonumber \\
&\quad& + \int_{\alpha_{\rm min}}^{\alpha_{\rm max}} \mathrm{d} \alpha \left\langle \sum_{i=1}^{N} | {\bf r}_i - {\bf r}_{0 i} |^2 \right\rangle_{\alpha}^{\rm T,S}.
\label{eq:S_alpha_min}
\end{eqnarray}

\subsubsection{Final expression of the glass entropy}

Finally, by combining Eqs.~(\ref{eq:S_glass_def}) and (\ref{eq:S_alpha_min}) we get the expression of $S_{\rm glass}$ as
\begin{eqnarray}
S_{\rm glass} &=&  \frac{Nd}{2} - N \ln \Lambda^d - \frac{Nd}{2} \ln \left(\frac{\alpha_{\rm max}}{\pi}\right)  \nonumber \\
&\quad& + N \lim_{\alpha_{\rm min} \to 0} \int_{\alpha_{\rm min}}^{\alpha_{\rm max}} \mathrm{d} \alpha \Delta_{\alpha}^{\rm T,S} + S_{\rm mix}^{(M)} -  \overline{ \mathcal{S}_{\rm mix}({\bf r}_0^N, \beta)}, \nonumber \\
\label{eq:S_glass_final}
\end{eqnarray}
where $\Delta_{\alpha}^{\rm T, S}$ is a mean-squared displacement defined by
\begin{equation}
\Delta_{\alpha}^{\rm T, S} = \frac{1}{N} \overline{\left\langle \sum_{i=1}^{N} | {\bf r}_i - {\bf r}_{0 i} |^2 \right\rangle_{\alpha}^{\rm T,S}},
\end{equation}
and $\mathcal{S}_{\rm mix}({\bf r}_0^N, \beta)$ is a mixing entropy contribution defined by
\begin{eqnarray}
\mathcal{S}_{\rm mix}({\bf r}_0^N, \beta) &=&  W({\bf r}_0^N, \beta)  - \beta U_0 ({\bf r}_0^N) \nonumber \\ 
&=& - \ln \left( \frac{1}{N!} \sum_{\pi} e^{-\beta \left( U_0(\Sigma_{\pi}^N, {\bf r}_0^N) - U_0({\bf r}_0^N) \right) } \right). \nonumber \\ 
\label{eq:s_mix_def_new}
\end{eqnarray}
In the derivation of Eq.~(\ref{eq:S_glass_final}) we also used the following relation: $\lim_{\alpha_{\rm min} \to 0} \beta \left\langle U_{\alpha_{\rm min}} (\Sigma_{\pi}^N, {\bf r}^N, {\bf r}_0^N) \right\rangle_{\alpha_{\rm min}}^{\rm T,S}  = \beta \left\langle U_0 (\Sigma_{\pi}^N, {\bf r}^N) \right\rangle_{0}^{\rm T,S} \nonumber 
= \beta \left\langle U_0 ({\bf r}^N) \right\rangle_{0}^{\rm T} \nonumber 
= \beta \overline{ U_0({\bf r}_0^N) }.
$

In Eq.~(\ref{eq:S_glass_final}), one can find two features that make our method distinct from the conventional Frenkel-Ladd method~\cite{frenkel1984new,coluzzi1999thermodynamics,sastry2000evaluation,angelani2007configurational}.
The first one is that the mean-squared displacement $\Delta_{\alpha}^{\rm T,S}$ has to be evaluated by Monte-Carlo simulations that sample both translational displacements and diameter swaps (as denoted by T, S). This should be distinguished from the normal mean-squared displacement $\Delta_{\alpha}^{\rm T}$ defined by using the average in Eq.~(\ref{eq:T}) instead of the one in Eq.~(\ref{eq:T_S}).
Due to the additional diameter swap moves, one expects that $\Delta_{\alpha}^{\rm T,S} \geq \Delta_{\alpha}^{\rm T}$ in general.
The second novel feature in Eq.~(\ref{eq:S_glass_final}) is the fact that $S_{\rm glass}$ contains a non-trivial mixing entropy term, $S_{\rm mix}^{(M)} - \overline{ \mathcal{S}_{\rm mix}}$.
For monodisperse particles or discrete mixtures where the swap of the diameters with different species have a high energy cost, the equalities, $\Delta_{\alpha}^{\rm T,S} = \Delta_{\alpha}^{\rm T}$ and $\overline{ \mathcal{S}_{\rm mix}}=S_{\rm mix}^{(M)}$ would hold, as we numerically confirm for a binary Lennard-Jones mixture.
In this case Eq.~(\ref{eq:S_glass_final}) reduces to the conventional Frenkel-Ladd method. On the other hand, for continuously polydisperse systems, one would expect $\Delta_{\alpha}^{\rm T,S} > \Delta_{\alpha}^{\rm T}$ and $\overline{ \mathcal{S}_{\rm mix}}/N < S_{\rm mix}^{(M=N)}/N \to \infty$. Therefore Eq.~(\ref{eq:S_glass_final}) is a straightforward generalization of the conventional Frenkel-Ladd method for systems with continuous polydispersity, and the thermodynamic integration automatically takes into account the correct number of permutations allowed by thermal fluctuations in equilibrium.  

The fact that $\Delta_{\alpha}^{\rm T,S} > \Delta_{\alpha}^{\rm T}$ also implies that $\Delta_{\alpha}^{\rm T,S}$ takes into account non-vibrational contributions due to the permutations of the diameters in addition to purely vibrational contribution measured by $\Delta_{\alpha}^{\rm T}$ (see related argument in Refs.~\onlinecite{ozawa2018ideal,ikeda2017mean}). Hence it is expected that the resulting $S_{\rm glass}$ more correctly deals with the non-vibrational contributions to the glass entropy as well.

\subsection{Computing the configurational entropy}

We summarize our computational scheme for the configurational entropy $S_{\rm conf}=S_{\rm tot}-S_{\rm glass}$. The entropies 
$S_{\rm tot}$ and $S_{\rm glass}$ are computed independently by two independent thermodynamic integrations. The entropy of the fluid
$S_{\rm tot}$ is obtained by the thermodynamic integration from the ideal gas, as described by Eqs.~(\ref{eq:total_soft}, \ref{eq:total_hard}), depending on the interaction potential. The glass entropy $S_{\rm glass}$ is obtained by a Frenkel-Ladd thermodynamic integration, summarized by Eq.~(\ref{eq:S_glass_final}).

It should be obvious, then, that the present scheme resolves the problem of an infinite mixing entropy for continuous polydispersity~\cite{ozawa2017does,baranau2017another}.
The diverging mixing entropy is the term $S_{\rm mix}^{(M)}$ in $S_{\rm tot}$ (through Eq.~(\ref{eq:entropy_ideal_gas})) which appears also in $S_{\rm glass}$ in Eq.~(\ref{eq:S_glass_final}). Instead $\overline{\mathcal{S}_{\rm mix}}$ in Eq.~(\ref{eq:S_glass_final}) remains as a finite mixing entropy contribution to $S_{\rm conf}$.
As we numerically confirm in Sec.~\ref{sec:numerical}, $\overline{\mathcal{S}_{\rm mix}}$ takes a finite value for continuously polydisperse systems, whereas it recovers the appropriate limit for discrete mixtures (see Appendix \ref{sec:meaning_s_mix}), and vanishes for monodisperse systems. 
Thus, the configurational entropy automatically incorporates the correct information about size polydispersity. Whereas the physical idea is the same as in Ref.~\onlinecite{ozawa2017does}, the present method is technically more elegant and does not require the approximate determination of a crossover in the evolution of the potential energy landscape. 

\section{Numerical implementation for three glass-formers}

\label{sec:numerical}

In this section, we numerically implement the method exposed in Sec.~\ref{sec:framework} for continuous polydisperse systems with soft and hard interactions, and for a standard binary Lennard-Jones mixture.
Since the results for $S_{\rm tot}$ can be found in the literature~\cite{berthier2017configurational,sastry2000evaluation}, we focus more specifically on the numerical determination of $S_{\rm glass}$. As seen in Eq.~(\ref{eq:S_glass_final}), the main computational tasks are the determination of the integral of $\Delta_{\alpha}^{\rm T, S}$ and the separate measurement of $\overline{\mathcal{S}_{\rm mix}}$. We illustrate these tasks separately for a single model, before presenting the final results for the three of them. 

\subsection{Models and simulation details}

We study three dimensional soft and hard sphere potential models using a continuous size polydispersity~\cite{ninarello2017models,berthier2016equilibrium}, 
where the particle diameter $\sigma$ of each particle is distributed from 
the following particle size distribution: 
$f(\sigma) = A\sigma^{-3}$, for $\sigma \in [ \sigma_{\rm min}, 
\sigma_{\rm max} ]$, choosing $\sigma_{\rm min} / \sigma_{\rm max} = 0.45$,
where $A$ is a normalization constant. 
We use the averaged diameter as the unit length. 
We simulate systems composed of $N$ particles in a cubic 
cell of volume $V$ with periodic boundary conditions~\cite{allen1989computer}. 

We use the following pairwise potential for a polydisperse soft sphere (SS) model~\cite{ninarello2017models},
\begin{eqnarray}
v_{ij}(r) &=& v_0 \left( \frac{\sigma_{ij}}{r} \right)^{12} + c_0 + c_1 \left( \frac{r}{\sigma_{ij}} \right)^2 + c_2 \left( \frac{r}{\sigma_{ij}} \right)^4, \label{eq:soft_v} \quad \\
\sigma_{ij} &=& \frac{(\sigma_i + \sigma_j)}{2} (1-\epsilon |\sigma_i - \sigma_j|), \label{eq:non_additive}
\end{eqnarray}
where $v_0$ is the unit of energy, and $\epsilon$  
quantifies the degree of non-additivity of the particle 
diameters. We set $\epsilon=0.2$.
The constants, $c_0$, $c_1$ and $c_2$, are chosen so that the first and 
second derivatives of $v_{ij}(r)$ become zero at the cut-off 
$r_{\rm cut}=1.25 \sigma_{ij}$.
We set the number density $\rho=N/V=1.0186$ with $N=1500$ for the soft 
sphere model.

For the polydisperse hard sphere (HS) model~\cite{berthier2016equilibrium}, we use the pair interaction which is zero for non-overlapping particles and infinite otherwise with the additive condition ($\epsilon=0$). 
However, we use a finite potential modeling of the hard sphere potential for $\overline{\mathcal{S}_{\rm mix}}$ (see Appendix~\ref{sec:s_mix_hard} for the details).
We perform the simulations for $N=1000$ and $8000$ to analyse finite-size effects. The hard sphere simulations are presented as a function of the 
reduced pressure $p=P/(\rho k_{\rm B} T)$, where $P$ is the measured 
pressure, and $k_{\rm B} T$ is set to unity. Thus, $1/p$ plays a role similar to the one of temperature for soft potentials. 

Finally, we study the standard Kob-Andersen (KA) binary Lennard-Jones model~\cite{kob1995testing}. Both species A and B have the same mass
and the concentration of each species are $X_{\rm A}=0.8$ and $X_{\rm B}=0.2$, respectively.  The interaction
potential between two particles is given by $v_{\alpha \beta}(r)
= 4\epsilon_{\alpha \beta}\{ (r/\sigma_{\alpha \beta} )^{12} -
(r/\sigma_{\alpha \beta})^{6}\}$, where $\alpha, \beta \in \{{\rm A},
{\rm B}\}$.  We set $\epsilon_{\rm AA}=1.0,\epsilon_{\rm AB}=1.5,\epsilon_{\rm BB}=0.5,
\sigma_{\rm AA}=1.0,  \sigma_{\rm AB}=0.8$ and $\sigma_{\rm BB}=0.88$.  The potential $v_{\alpha
\beta}(r)$ is truncated and shifted at $r_{\rm cut}=2.5\sigma_{\alpha \beta}$.
We show energy in units of $\epsilon_{\rm AA}$, with the Boltzmann constant
$k_{\rm B}=1$, and length in units of $\sigma_{\rm AA}$. Simulations are performed at constant
density $\rho=1.2$.  The number of particles is $N=1200$.

We prepare equilibrium configurations for continuously polydisperse systems using swap Monte-Carlo simulations~\cite{ninarello2017models,berthier2016equilibrium}. With probability $P_{\rm swap}=0.2$ we perform a swap move where we pick two particles at random and attempt to exchange their diameters and with probability $1-P_{\rm swap}=0.8$, we perform conventional Monte Carlo translational moves. 
Equilibrium configurations for the KA model are prepared using standard Monte Carlo simulations~\cite{berthier2007monte} (i.e., without swap moves, $P_{\rm swap}=0$). Lower temperature configurations of the KA model are prepared by the parallel tempering algorithm~\cite{marinari1992simulated,hukushima1996exchange} produced in Ref.~\onlinecite{coslovich2018dynamic}.
The statistical averages shown in Eqs.~(\ref{eq:T_S}), (\ref{eq:T}), and (\ref{eq:S}) are performed by using $P_{\rm swap}=0.2$, $0.0$, and $1.0$ for Eqs.~(\ref{eq:T_S}), (\ref{eq:T}), and (\ref{eq:S}), respectively.
The statistical average in Eq.~(\ref{eq:template_average}) is performed by averaging over 5-20 independent reference configurations.

To present results for the three models coherently, we use a temperature $T^*$ normalized by the mode coupling crossover.
We define $T^*=T/T_{\rm mct}$ for the polydisperse soft spheres ($T_{\rm mct}=0.104$)~\cite{ninarello2017models} and the Kob-Andersen model ($T_{\rm mct}=0.435$)~\cite{kob1995testing}.
For the polydisperse hard spheres, we define $T^*=p_{\rm mct}/p$ with $p_{\rm mct}=23.5$~\cite{berthier2017configurational}.

\subsection{Constrained mean-squared displacements}

\begin{figure}
\includegraphics[width=0.95\columnwidth]{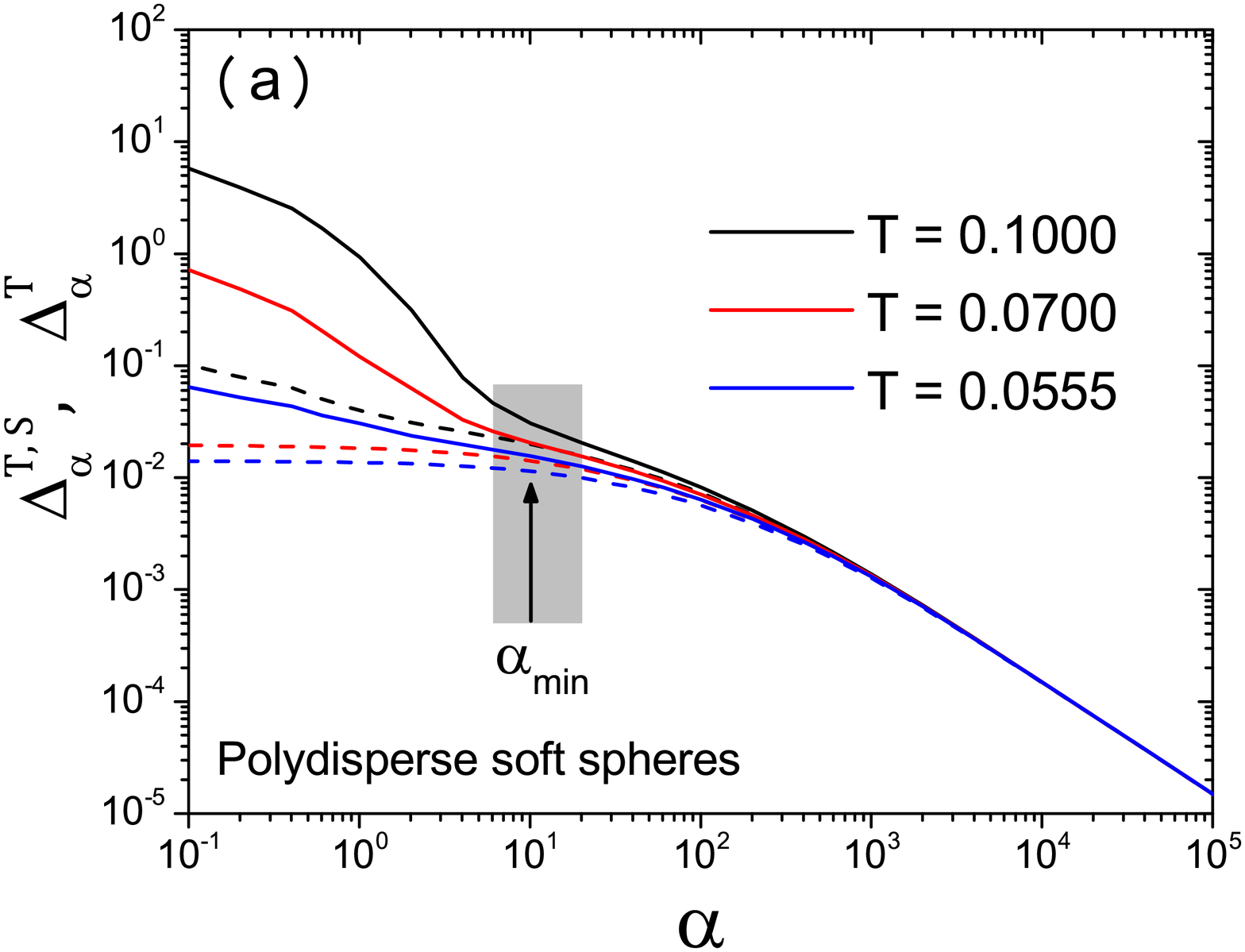}
\includegraphics[width=0.95\columnwidth]{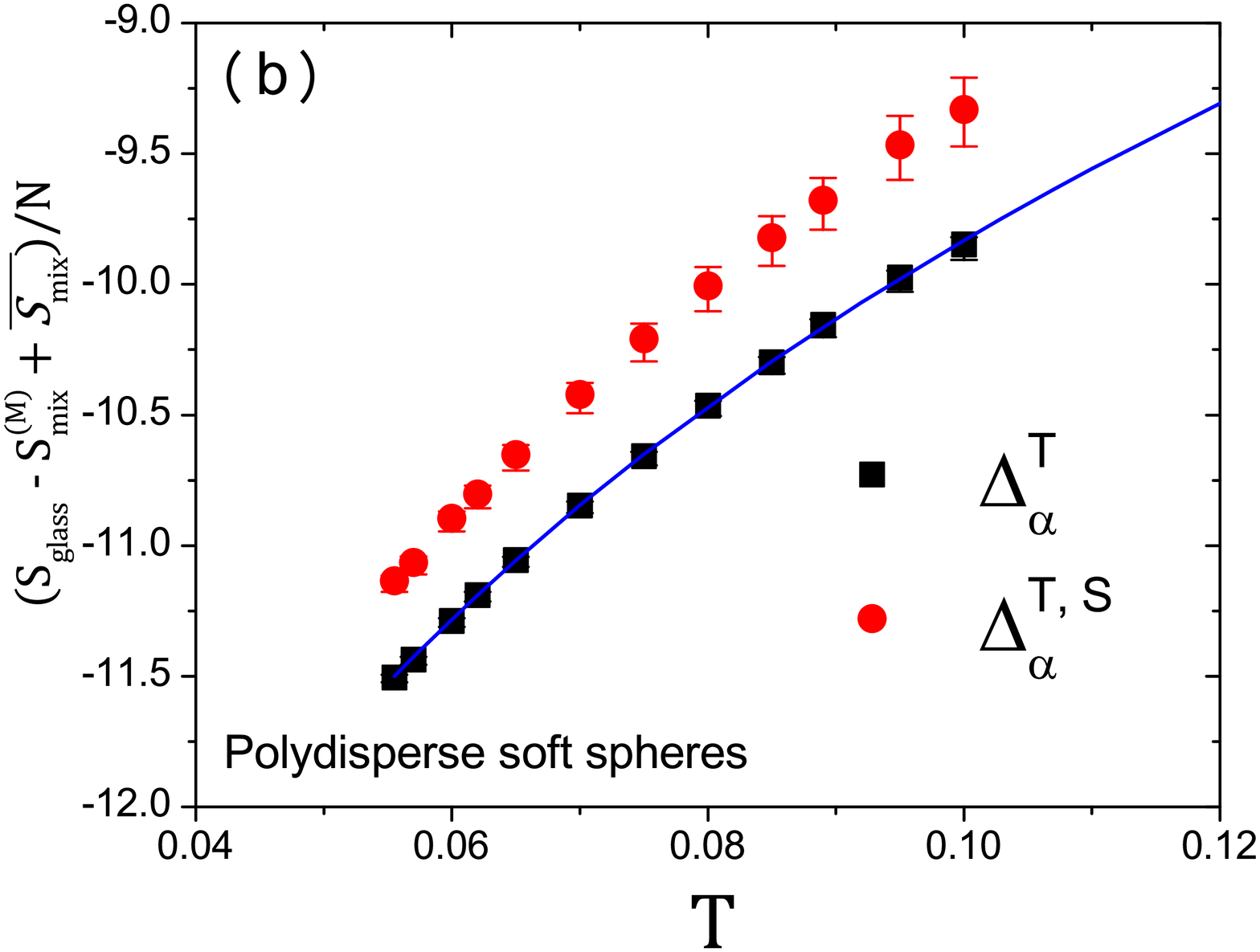}
\caption{(a) Mean-squared displacement in the Frenkel-Ladd construction with normal ($\Delta_{\alpha}^{\rm T}$: dashed-line) and diameter swap ($\Delta_{\alpha}^{\rm T, S}$: solid-line) Monte-Carlo simulations for polydisperse soft spheres. The shaded region corresponds to $\alpha_{\rm min} \in [6.1, 20.2]$ and the arrow indicates $\alpha_{\rm min}=10.1$.
(b) Glass entropy $S_{\rm glass}/N$ obtained by Eq.~(\ref{eq:S_glass_final}) using either $\Delta_{\alpha}^{\rm T, S}$ and $\Delta_{\alpha}^{\rm T}$ for $\alpha_{\rm min}=10.1$. The mixing entropy terms $S_{\rm mix}^{(M)} - \overline{\mathcal{S}_{\rm mix}}$ are subtracted from $S_{\rm glass}$. The errorbars correspond to $S_{\rm glass}/N$ computed in the region $\alpha_{\rm min} \in [6.1, 20.2]$. The full blue line is the vibrational entropy $S_{\rm vib}/N=(S_{\rm harm}+S_{\rm anh})/N$, where $S_{\rm harm}$ and $S_{\rm anh}$ are obtained by diagonalization of the Hessian matrix in the inherent structure and its anharmonic correction, respectively~\cite{berthier2017configurational}.}
\label{fig:FL}
\end{figure}  

In this section, we illustrate the numerical determination of the integral of $\Delta_{\alpha}^{\rm T, S}$ which appears in Eq.~(\ref{eq:S_glass_final}). 
Starting from $\alpha=\alpha_{\rm max}=3.0 \times 10^6 - 1.01 \times 10^7$ (see below), we perform MC simulations with decreasing $\alpha$ in steps of $\delta ({\rm log}_{10} \alpha) \simeq 0.18-0.4$.
For each data point, we perform $\tau = 2 \times 10^4 - 2 \times 10^6$ MC steps, measuring $\Delta_{\alpha}^{\rm T, S}$ only in the second half of the simulation.
In Fig.~\ref{fig:FL}(a) we show the evolution of $\Delta_{\alpha}^{\rm T, S}$ with the strength of the harmonic coupling $\alpha$, for polydisperse soft spheres at several temperatures. As expected, $\Delta_{\alpha}^{\rm T, S}$ is very small at large $\alpha$ and increases as $\alpha$ decreases. 
When obtaining the data at various values of $\alpha$ we have to make sure that the mean-squared displacements have converged to the correct equilibrium value. We have performed detailed numerical tests for this convergence. We have measured $\Delta_{\alpha}^{\rm T, S}$ by changing the timescale $\tau$ over which $\alpha$ is varied and confirmed that $\Delta_{\alpha}^{\rm T, S}$ does not depend on $\tau$ down to $\alpha_{\rm min}$ chosen in this study (see below). 
We also applied tests where $\Delta_{\alpha}^{\rm T, S}$ is measured starting from both the reference configuration and from an annealed configuration produced by the swap MC simulation at higher temperature.
The two simulations provide consistent results, ensuring the equilibration. These tests show that it is much easier to converge constrained simulations in the Frenkel-Ladd setup than in any other scheme (such as cavity measurements~\cite{biroli2008thermodynamic}). This is consistent with the results of Ref.~\onlinecite{berthier2012static}, which already showed that cavity measurements were the most difficult constrained scheme to obtain equilibrium measurements. A possible explanation of this qualitative difference is that a (soft) constrained is locally applied to each particle in the Frenkel-Ladd method, whereas a (hard) global constraint is applied from the boundary in cavity measurements.   

To understand the effect of the particle diameter permutations on the measured cage, we also show the evolution of $\Delta_{\alpha}^{\rm T}$ for the same temperatures with dashed lines. The two mean-squared displacements then only differ by the introduction in $\Delta_{\alpha}^{\rm T, S}$ of particle diameter permutations. 

For strong $\alpha$, both $\Delta_{\alpha}^{\rm T, S}$ and $\Delta_{\alpha}^{\rm T}$ precisely obey the Einstein solid prediction, $\Delta_{\alpha}^{\rm T, S} \approx \Delta_{\alpha}^{\rm T} \approx 3/(2\alpha)$. 
With decreasing $\alpha$, $\Delta_{\alpha}^{\rm T, S}$ and $\Delta_{\alpha}^{\rm T}$ enter a plateau region shown by the shaded region.
In this region, the system is trapped by its own cage.
We find that $\Delta_{\alpha}^{\rm T, S} > \Delta_{\alpha}^{\rm T}$, which means that $\Delta_{\alpha}^{\rm T, S}$ samples a larger phase space within the glass state than $\Delta_{\alpha}^{\rm T}$.
Decreasing $\alpha$ further, the harmonic constraint for $\Delta_{\alpha}^{\rm T, S}$ is too weak and the metastability of the glass state is not strong enough to prevent the system from diffusing, which translates into an upturn of $\Delta_{\alpha}^{\rm T, S}$ for higher temperature at small $\alpha$. The effect is also visible for $\Delta_{\alpha}^{\rm T}$, but it is much less pronounced since the structural relaxation without swap moves is considerably slower~\cite{ninarello2017models}, and metastability is therefore stronger.

To perform the integration and to take the $\alpha_{\rm min} \to 0$ limit in Eq.~(\ref{eq:S_glass_final}), we use the following manipulation:
\begin{equation}
\lim_{\alpha_{\rm min} \to 0} \int_{\alpha_{\rm min}}^{\alpha_{\rm max}} \mathrm{d} \alpha \Delta_{\alpha}^{\rm T,S}  \simeq \alpha_{\rm min}  \Delta_{\alpha_{\rm min}}^{\rm T,S}  + \int_{\alpha_{\rm min}}^{\alpha_{\rm max}} \mathrm{d} \alpha \Delta_{\alpha}^{\rm T,S}.
\label{eq:integration_Delta}
\end{equation}
The practical choice for $\alpha_{\rm max}$ is simple, as it is sufficient that it lies deep inside the Einstein solid regime. We choose $\alpha_{\rm max}=3.0 \times 10^6 - 1.01 \times 10^7$ for all systems. We set $\alpha_{\rm min}=10.1$ for the polydisperse soft spheres within the plateau region indicated in the arrow in Fig.~\ref{fig:FL}(a), where the equilibration is ensured.

We show the resulting glass entropy minus the mixing entropy contribution, $(S_{\rm glass} - S_{\rm mix}^{(M)} + \overline{\mathcal{S}_{\rm mix}})/N$, in Fig.~\ref{fig:FL}(b). (The mixing entropy terms are considered in the following subsection.) We also present the results obtained by substituting 
$\Delta_{\alpha}^{\rm T,S}$ by $\Delta_{\alpha}^{\rm T}$ in Eq.~(\ref{eq:integration_Delta}) to get some feeling about the quantitative importance of particle diameter permutations in this measurement. We also compare the value of the same glass entropy contribution obtained by following the potential energy landscape recipe~\cite{sciortino2005potential}, where a vibrational entropy $S_{\rm vib}$ is computed as $S_{\rm vib}=S_{\rm harm}+S_{\rm anh}$, where $S_{\rm harm}$ and $S_{\rm anh}$ are the entropies obtained by diagonalization of the Hessian matrix at the inherent structure and its anharmonic correction, respectively~\cite{berthier2017configurational}.
 
Strikingly, we find that the glass entropy obtained by the ordinary Frenkel-Ladd approach with no diameter permutation takes values very similar to the vibrational entropy $S_{\rm vib}$ computed by the potential energy landscape approach. This trend suggests that $\Delta_{\alpha}^{\rm T}$ accounts for purely vibrational motion inside a single inherent structure~\cite{ozawa2018ideal}. We also find the same trend in the KA model (not shown). On the other hand, the glass entropy $S_{\rm glass}$ obtained with diameter permutation using $\Delta_{\alpha}^{\rm T, S}$ takes larger values, because $\Delta_{\alpha}^{\rm T, S} > \Delta_{\alpha}^{\rm T}$. In other words, $S_{\rm glass}$ takes into account non-vibrational contributions, which should be associated with the presence of many inherent structures within a single glass state~\cite{birolimonasson,ozawa2017does,ozawa2018ideal}. The association of many inherent structures within a single glass state is impossible within the potential energy landscape and ordinary Frenkel-Ladd approaches, but arises naturally within both the present scheme and the Franz-Parisi free-energy measurement~\cite{berthier2014novel}.

Note that the specific choice of the value of $\alpha_{\rm min}$ mostly affects the determination of $S_{\rm glass}$ at higher temperature, where the plateau is not well formed. To estimate this effect, we draw errorbars whose range corresponds to $S_{\rm glass}/N$ obtained from the edges of the shaded region, $\alpha_{\rm min} \in [6.1, 20.2]$,  in Fig.~\ref{fig:FL}(b).
We find that the size of the errorbars progressively becomes smaller as the temperature decreases, in agreement with the clear plateau formation at the lower temperature in Fig.~\ref{fig:FL}(a). This trend justifies our choice of $\alpha_{\rm min}$ at low temperatures.

We find qualitatively similar behavior for the polydisperse hard sphere model and the KA model (not shown). However, whereas the inequality $\Delta_{\alpha}^{\rm T, S} > \Delta_{\alpha}^{\rm T}$ holds for the polydisperse hard sphere model similarly to the soft sphere model, the KA model shows $\Delta_{\alpha}^{\rm T, S} \approx \Delta_{\alpha}^{\rm T}$ due to the fact that diameter permutations are hardly accepted in this bidisperse model~\cite{flenner2006hybrid}. 

\subsection{Mixing entropy}

To measure $\mathcal{S}_{\rm mix}({\bf r}_0^N, \beta)$ numerically, we perform a thermodynamic integration over a temperature $\beta'$ from the target temperature $\beta'=\beta$ with a given reference configuration ${\bf r}_0^N$ to the high temperature limit, $\beta' \to 0$.
The high temperature limit of Eq.~(\ref{eq:s_mix_def_new}) is trivially
$\mathcal{S}_{\rm mix}({\bf r}_0^N, \beta' \to 0) \to 0$.
The derivative of $\mathcal{S}_{\rm mix}({\bf r}_0^N, \beta')$ with respect to $\beta'$ becomes a potential energy difference,
$\frac{\partial \mathcal{S}_{\rm mix} ({\bf r}_0^N, \beta') }{\partial \beta'} =  \left\langle U_0(\Sigma_{\pi}^N, {\bf r}_0^N) \right\rangle_{\beta'}^{\rm S} - U_0({\bf r}_0^N) \equiv  \Delta U_{\rm mix}({\bf r}_0^N, \beta')$.
In this last expression, $\Delta U_{\rm mix}$ quantifies the potential energy increment due to the exploration of the permutation phase space by heating the system at temperature $T' = 1/\beta'> T$. Therefore, we get by thermodynamic integration,
\begin{equation}
\overline{\mathcal{S}_{\rm mix}({\bf r}_0^N, \beta)} =  \int_0^{\beta} \mathrm{d} \beta' \overline{\Delta U_{\rm mix}({\bf r}_0^N, \beta')}.
\label{eq:s_mix_integral}
\end{equation}
To measure $\Delta U_{\rm mix}({\bf r}_0^N, \beta')$ in practice, the system is gradually heated from the target temperature $\beta'=\beta$ to the infinite temperature $\beta' \to 0$ by performing Monte Carlo simulations where only particle diameter permutations are attempted (denoted by the superscript `S' in Eq.~(\ref{eq:S})) while keeping fixed the particle positions of the reference configuration ${\bf r}_0^N$ generated at $\beta$.

\begin{figure}
\includegraphics[width=0.95\columnwidth]{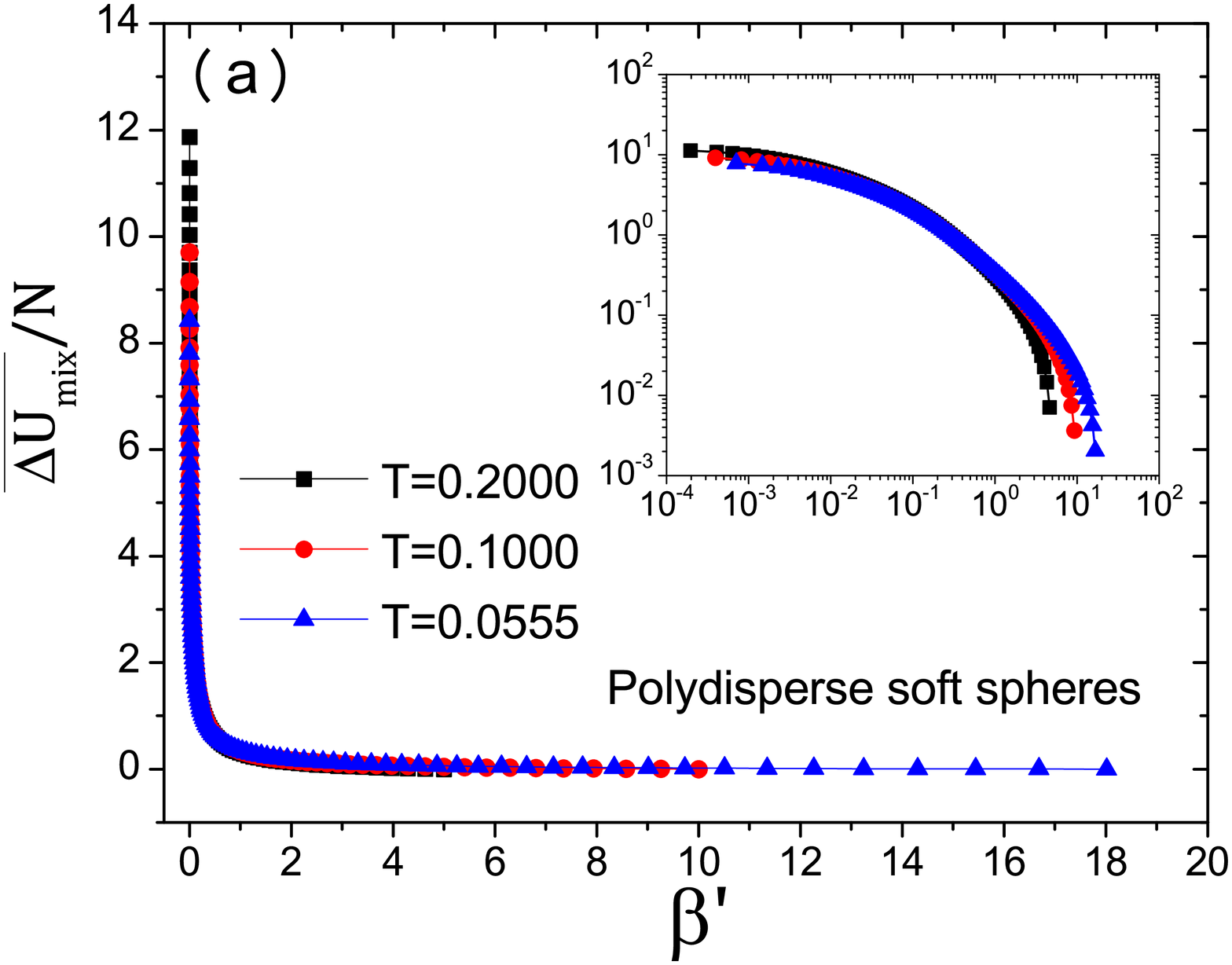}
\includegraphics[width=0.95\columnwidth]{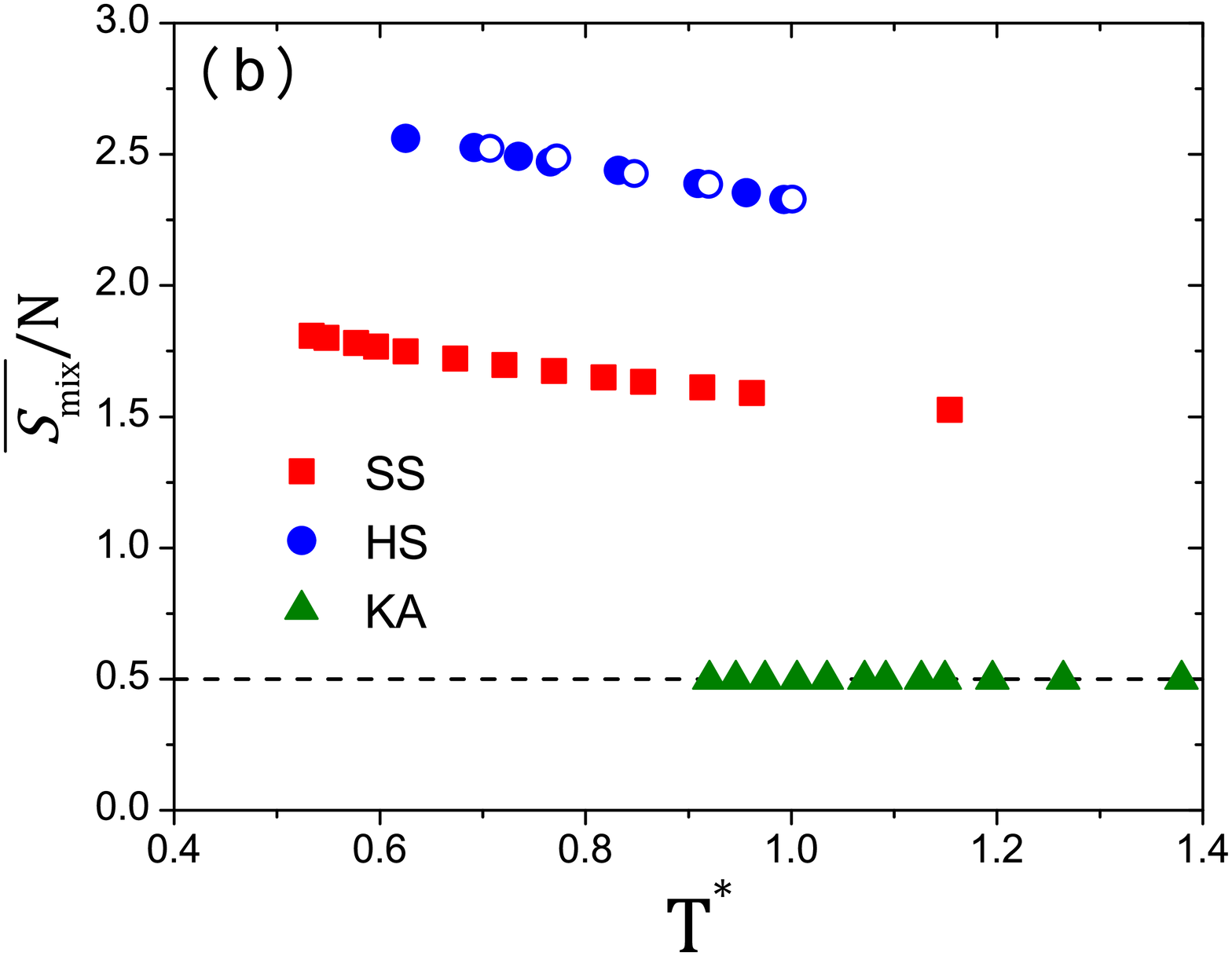}
\caption{(a) Evolution of $\overline{\Delta U_{\rm mix}({\bf r}_0^N, \beta')}$ during the thermodynamic integration over $\beta'$ for several reference temperatures $\beta$. Inset: same data in log-log representation.  
(b) Mixing entropy $\overline{\mathcal{S}_{\rm mix}}/N$ obtained by Eq.~(\ref{eq:s_mix_integral}) as a function of the normalized 
temperature $T^* = T/T_{\rm mct}$ for the three studied models.
Filled and empty circles for HS correspond to $N=1000$ and $8000$, respectively. The dashed-line corresponds to the combinatorial mixing entropy for the KA mixture.}
\label{fig:s_mix}
\end{figure}  

As shown in Fig.~\ref{fig:s_mix}(a) for polydisperse soft spheres, $\overline{\Delta U_{\rm mix}}/N$ takes a very small value at large $\beta'$, and sharply increases approaching $\beta' \to 0$. This is observed for all temperatures $T=1/\beta$, with a relatively weak temperature dependence.
Note that $\overline{\Delta U_{\rm mix}}/N$ remains finite  as $\beta' \to 0$, as shown in the inset. This guarantees a finite mixing entropy $\overline{\mathcal{S}_{\rm mix}}/N$ as well. A
qualitatively similar behavior is found for polydisperse hard spheres and for the KA model, except that the KA model shows fully temperature-independent results. To compute  $\overline{\Delta U_{\rm mix}}$ for the hard spheres, we use a soft potential modeling, as described in Appendix~\ref{sec:s_mix_hard}.
We also perform a cooling path from $\beta' = 0$ to $\beta'=\beta$ for polydisperse soft spheres, which coincides perfectly with the heating path described above. Therefore, we conclude that one can easily achieve an equilibrium path for the thermodynamic integration and sample the permutation-phase space properly.

In Fig.~\ref{fig:s_mix}(b) we show the resulting $\overline{\mathcal{S}_{\rm mix}}/N$ as a function of the normalized temperature $T^*$ for the three studied systems. For the KA model, $\overline{\mathcal{S}_{\rm mix}}/N$ precisely recovers the standard combinatorial mixing entropy $S_{\rm mix}^{(M=2)}/N =- X_{\rm A} \ln X_{\rm A} - X_{\rm B} \ln X_{\rm B} \simeq 0.5$ (with $X_{\rm A}=0.8$ and $X_{\rm B}$=0.2) for a wide range of temperatures. This means that $\overline{\mathcal{S}_{\rm mix}}=S_{\rm mix}^{(M=2)}$ holds and that the mixing entropy terms in Eq.~(\ref{eq:S_glass_final}) exactly cancel each other, directly justifying previous treatments of the mixing entropy for this model~\cite{sciortino1999inherent,sastry2000evaluation}.
We find that this treatment holds in binary hard sphere mixtures with sufficiently large size ratio as well, as demonstrated in Appendix~\ref{sec:meaning_s_mix}. We also find that $\overline{\mathcal{S}_{\rm mix}}/N$ smoothly connects the monodisperse limit where $\overline{\mathcal{S}_{\rm mix}}/N=0$ to the large size ratio regime where $\overline{\mathcal{S}_{\rm mix}}=S_{\rm mix}^{(M=2)}$, as shown in Appendix~\ref{sec:meaning_s_mix}. These results mean that we do not need to {\it decide} how to treat the system (as being monodisperse or polydisperse~\cite{coluzzi1999thermodynamics}) since our method directly measures the correct value of the mixing entropy. This is conceptually analogous to a recent analytic computation~\cite{ikeda2016note}, although our approach can deal with a continuous polydispersity more straightforwardly.  

The important result is of course that for the continuously polydisperse systems, $\overline{\mathcal{S}_{\rm mix}}/N$ takes slightly larger values, but it remains finite. The obtained values are comparable to our previous estimates through an effective $M^*$-component approximation~\cite{ozawa2017does}.
In this description, $M^*$ was obtained by dividing the particle diameter distribution $f(\sigma)$ into a series of $M^*$ finite intervals of the same width, $\Delta \sigma=(\sigma_{\rm max}-\sigma_{\rm min})/M^*$.
Interestingly, however, we find that $\overline{\mathcal{S}_{\rm mix}}$ slightly increases with decreasing the temperature or increasing the pressure, an effect that was not captured by the previous estimation. To obtain a more quantitative comparison with our previous work, we may consider the quantity  $M^\dagger=\exp[\overline{\mathcal{S}_{\rm mix}}/N]$ which can be seen as an effective number of components for the system using the assumption of equal concentrations, i.e., $X_m=1/M^\dagger$ ($m=1, 2, \cdots, M^\dagger$).
As a consequence of the slight increase of $\overline{\mathcal{S}_{\rm mix}}$, $M^\dagger$ also increases steadily with decreasing the temperature or increasing the pressure, which means that a smaller $\Delta \sigma$ is  effectively needed to properly represent the continuous mixture with increasing the degree of supercooling. The range of $M^\dagger$ in Fig.~\ref{fig:s_mix}(b) is $M^\dagger \simeq 5 - 6$ for polydisperse soft spheres, and $M^\dagger \simeq 10 - 13$ for polydisperse hard spheres.
These results suggest that the hard sphere potential is more sensitive to small diameter differences than the soft potential. 

Note finally that our measurement of $\overline{\mathcal{S}_{\rm mix}}$ is not influenced by finite size effects, as can be seen by comparing $N=1000$ and $N=8000$ data for hard spheres in Fig.~\ref{fig:s_mix}(b).

\subsection{Configurational entropy for three glass-formers}

\begin{figure}
\includegraphics[width=0.95\columnwidth]{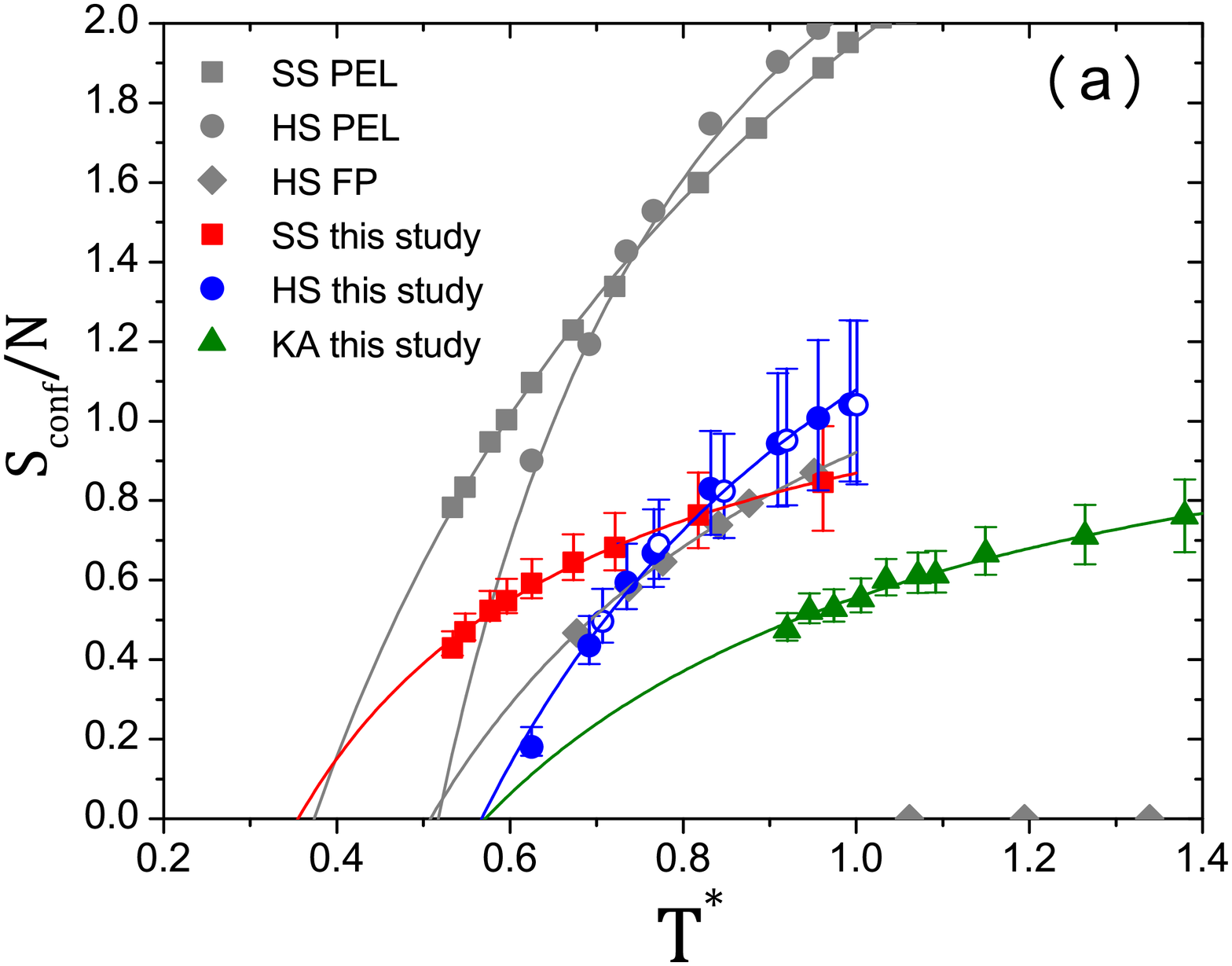}
\includegraphics[width=0.95\columnwidth]{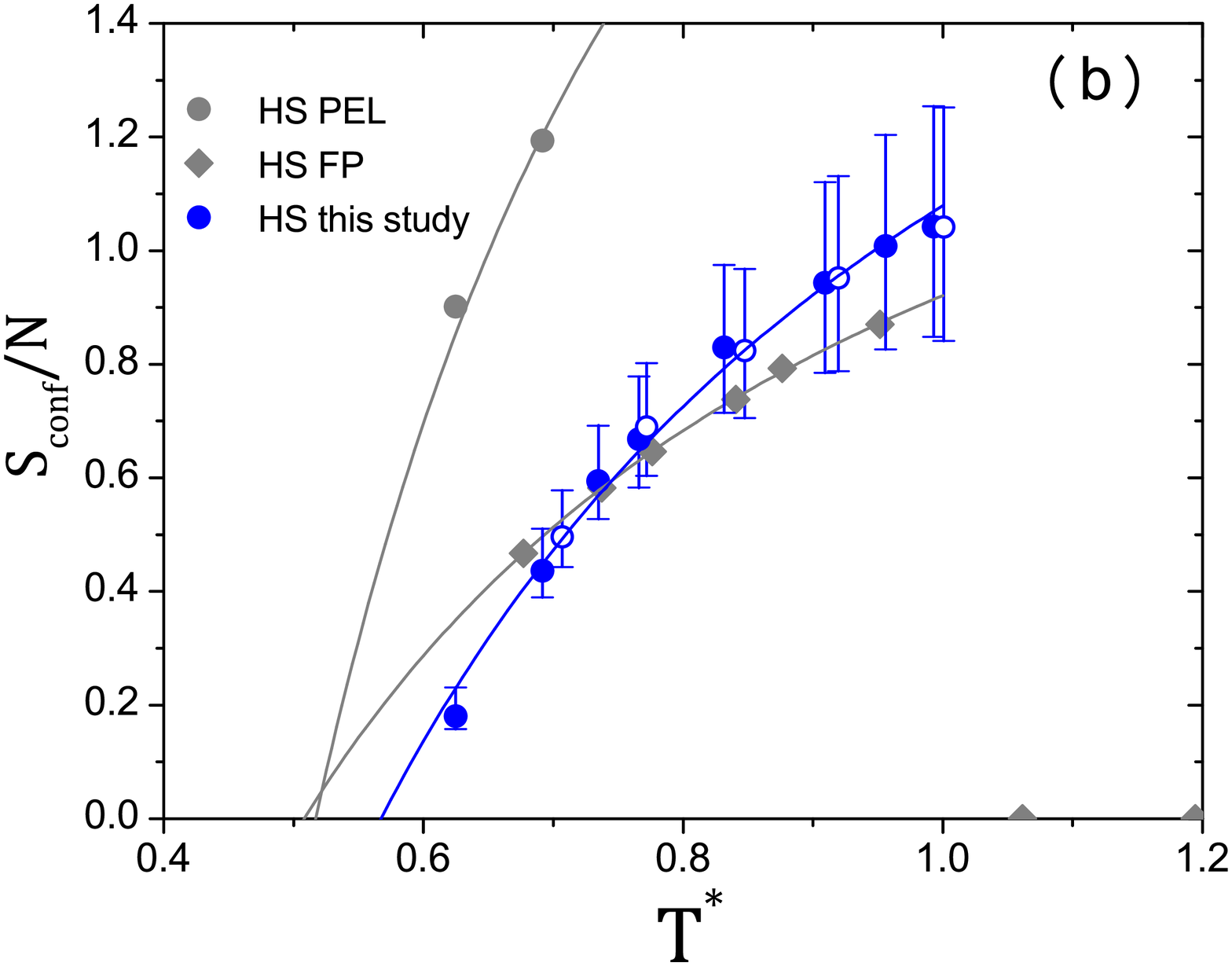}
\caption{
(a) Configurational entropy $S_{\rm conf}/N$ obtained for three glass-formers with errorbars reflecting the chosen range of $\alpha_{\rm min}$. 
Filled and empty circles for hard spheres (HS) correspond to $N=1000$ and $8000$, respectively.
$S_{\rm conf}/N$ based on the potential energy landscape (PEL) approach for soft spheres (SS) and HS in Ref.~\onlinecite{berthier2017configurational} are plotted using grey symbols.  
The Franz-Parisi (FP) potential approach for HS is also shown.
Extrapolations are performed by fitting data to $S_{\rm conf}/N=A(1-T_{\rm K}^*/T^*)$ or using the curves in Ref.~\onlinecite{berthier2017configurational}. 
(b) Zoom of the low temperature data for HS.}
\label{fig:s_conf}
\end{figure}  

Finally, we compile the configurational entropy, $S_{\rm conf}/N=(S_{\rm tot}-S_{\rm glass})/N$, of three systems as a function of the normalized temperature $T^*$ in Fig.~\ref{fig:s_conf}. Since 
$S_{\rm conf}$ depends on the chosen $\alpha_{\rm min}$ in the determination of $S_{\rm glass}$, we display the errorbars corresponding to $S_{\rm conf}$ from $\alpha_{\rm min}$-values chosen inside the plateau region, in the same way as in Fig.~\ref{fig:FL}(b). The size of the errorbars decreases with decreasing $T^*$ for all systems, showing a systematic improvement of the accuracy of our measurement towards lower temperature. 
The range of chosen $\alpha_{\rm min}$ are $\alpha_{\rm min}=10.1$, $\alpha_{\rm min} \in [6.1, 20.2]$ for polydisperse soft spheres, $\alpha_{\rm min}=15.1$, $\alpha_{\rm min} \in [7.5, 30.1]$ for polydisperse hard spheres, and $\alpha_{\rm min}=10.0$, $\alpha_{\rm min} \in [4.0, 20.0]$ for the Kob-Andersen model, respectively. We also find that our measurements of $S_{\rm conf}$ do not involve finite size effects, as shown by the comparison between $N=1000$ and $N=8000$ for hard spheres.

To extrapolate $S_{\rm conf}$ down to lower temperatures, we use an empirical relation, $S_{\rm conf}/N=A(1-T_{\rm K}^*/T^*)$, where $A$ and $T_{\rm K}^*$ are fitting parameters~\cite{richert1998dynamics,banerjee2014role}.
The numerical results of all models suggest that $S_{\rm conf}/N$ vanishes at a finite $T_{\rm K}^*>0$, which  consolidates previous findings~\cite{berthier2017configurational}. Specifically, we find 
$T_{\rm K}^* = 0.355, 0.567$ and $0.571$ for soft spheres, hard spheres, and the KA model, respectively. 
However, it is clear from the data shown in Fig.~\ref{fig:s_conf} that the possibility that a sharp Kauzmann transition is eventually avoided is also compatible with our data, if some presently-inaccessible crossover temperature exists below which the temperature evolution of the configurational entropy changes qualitatively, as envisioned in several analytical models~\cite{debenedetti2003model,tarjus2005frustration,effort3}.

We plot other estimates of $S_{\rm conf}/N$ obtained in Ref.~\onlinecite{berthier2017configurational}, shown as squares (polydisperse soft spheres) and circles (polydisperse hard spheres).
These estimates are based on the potential energy landscape description of $S_{\rm glass}$~\cite{sciortino2005potential} together with a combinatorial approximation of the mixing entropy using the effective $M^*$-components approximation~\cite{ozawa2017does}. We also plot $S_{\rm conf}$ obtained by the Franz-Parisi free energy~\cite{franz1997phase,berthier2014novel} for polydisperse hard spheres. 
We find that $S_{\rm conf}/N$ by our scheme for the polydisperse systems take smaller values than those of the PEL approach, mainly due to the fact that non-vibrational contributions are more correctly taken into account~\cite{ozawa2018ideal}. However, overall, the estimated Kauzmann temperatures $T_{\rm K}^*$ are quite  consistent among the different measurements of $S_{\rm conf}$. 

Remarkably, our new scheme produces values that are comparable to $S_{\rm conf}$ obtained from the Franz-Parisi free energy~\cite{berthier2014novel} for polydisperse hard spheres, as highlighted in Fig.~\ref{fig:s_conf}(b). Our numerical results imply that these two methods seemingly sample similar regions of the free-energy landscape. We find however a slight difference of the functional form and the resulting location of $T_{\rm K}^*$. We note that choosing a state point dependent $\alpha_{\rm min}$ for our scheme might slightly change the functional form inside the range of the errorbar.
Similarly, the definition of the overlap function in the Franz-Parisi potential and the choice of a coarse-graining length would also affect the detailed functional form of these results. 

We emphasize that the main difference between these two estimates does not simply originate from computational details, since the physical construction is qualitatively different between the two approaches. In the present scheme, we use Eq.~(\ref{eq:def}) to separately compute the fluid entropy $S_{\rm tot}$ (by thermodynamic integration from the ideal gas) and the glass entropy $S_{\rm glass}$ (from thermodynamic integration from an `ideal' Einstein solid). Each integration is relatively straighforward as it does not involve the crossing any equilibrium phase transition since the fluid and solid phases are treated separately. Instead, the Franz-Parisi free energy provides $S_{\rm conf}$ in a single measurement, by following an equilibrium path from the equilibrated fluid up to the glass state confined in a configuration space. This path however involves crossing an equilibrium phase transition~\cite{berthier2013,berthier2014novel,BJ15}, and it is therefore computationally more costly. Of course, ideally these two methods should be able to produce consistent results.  

\section{Discussion and conclusion}

\label{sec:conclusion}

We have developed a computational scheme to measure the configurational entropy for generic polydisperse systems, which is a straightforward generalization of the conventional Frenkel-Ladd approach. The key idea is the introduction of diameter permutations as additional degrees of freedom for the glass entropy, which is implemented by a simple swap Monte Carlo algorithm. 
Our scheme automatically takes into account the mixing entropy contribution for any particle size distribution as well as non-vibrational contributions to the glass entropy. This provides an accurate configurational entropy determination which seems comparable to the free energy approach based on the Franz-Parisi potential. This is quite remarkable because the physical construction in the two approaches are qualitatively different.
A practical merit of our method is a relatively low computational cost, which allows us to study more deeply supercooled and larger systems.
There is still a slight discrepancy of the functional forms between our scheme and the Franz-Parisi free energy, which might be cured by more precise choices for $\alpha_{\rm min}$ and for the definition of the overlap function.
Furthermore, the distinction between the two methods is still quite large in the Kob-Andersen model~\cite{sciortino1999inherent,BJ15}. Consolidating the mutual consistency among different configurational entropy measurements would be an important step for the complete thermodynamic characterization of the nature of the glass transition~\cite{2dglass}.

It has been argued that the entropy of colloidal polydisperse systems involves a subjective measurement, because particle distinguishability depends on the resolution chosen by the observer~\cite{cates2015celebrating,frenkel2014colloidal}.
This argument seems to prohibit a well-defined and quantitative value of the configurational entropy for colloidal glasses. However, our proposed scheme is free from any conceptual and technical difficulties due to continuous polydispersity thanks to a proper statistical mechanics description of the glass state. Thus, the observer subjectivity plays no role in our measurement.
Note that outside the realm of the configurational entropy measurement discussed here, the entropy of colloidal systems in the fluid still remains plagued with potential infinity problems, which should be managed for each case separately~\cite{warren1998combinatorial,swendsen2006statistical,maynar2011entropy,frenkel2014colloidal,paillusson2014role,cates2015celebrating,paillusson2018gibbs}.
Among them, our scheme might be useful also for phase equilibria problems in the canonical ensemble~\cite{sollich2001predicting,wilding2010phase} or accurate determination of the entropy of granular materials~\cite{asenjo2014numerical,martiniani2017numerical}.

\begin{acknowledgments}

We thank H. Ikeda, A. Ninarello, and F. Zamponi for helpful discussions. We warmly thank W. Kob and A. Ninarello for sharing very low temperature configurations.
This work is supported by a grant from the Simons Foundation (No. 454933, LB, No. 454949, GP).
\end{acknowledgments}

\appendix

\section{$\mathcal{S}_{\rm mix}$ for simple mixtures}
\label{sec:meaning_s_mix}

We demonstrate that $\mathcal{S}_{\rm mix}$ reduces to the standard combinatorial mixing entropy $S_{\rm mix}^{(M)}$ for simple mixtures.

\subsection{Monodisperse and binary mixtures}

First, it is instructive to verify that $\mathcal{S}_{\rm mix}$ vanishes in the monodisperse limit.
In this limit, since $U_0(\Sigma_{\pi}^N, {\bf r}_0^N)=U_0({\bf r}_0^N)$ for any permuation $\Sigma_{\pi}^N$, we immediately get from Eq.~(\ref{eq:s_mix_def_new})
that $\mathcal{S}_{\rm mix} ({\bf r}_0^N, \beta) =  0$.

Next, we consider the case of $M=2$ binary mixtures composed of species A and B with concentrations $X_{\rm A}=N_{\rm A}/N$ and $X_{\rm B}=N_{\rm B}/N$ ($0< X_{\rm A}, X_{\rm B} < 1$).
Starting from a reference equilibrium configuration ${\bf r}_0^N$ with a potential energy $U_0 ({\bf r}_0^N)=U_0 (\Sigma_{\pi^*}^N, {\bf r}_0^N)$,
the system may explore different permutations $\Sigma_{\pi}^N$.
Permutations associated with the exchange of diameters within the same species (denoted by A $\leftrightarrow$ A or B $\leftrightarrow$ B) have a  strictly zero energy cost. There exist $N_{\rm A}! N_{\rm B}!$ such permutations.
On the other hand, at sufficiently low temperature or high density, permutations associated with an exchange of the diameters between different species (denoted by A $\leftrightarrow$ B) may produce a high energy cost.
Therefore, we can evaluate the Bolzmann factor in that case as
\begin{equation}
e^{-\beta \left( U_0(\Sigma_{\pi}^N, {\bf r}_0^N) - U_0({\bf r}_0^N) \right) } \simeq \left \{ \begin{array}{ll}
    1 & (\Sigma_{\pi}^N \ {\rm with} \ {\rm A} \leftrightarrow {\rm A} \ {\rm or} \ {\rm B} \leftrightarrow {\rm B}), \\
    0 & (\Sigma_{\pi}^N \ {\rm with} \ {\rm A} \leftrightarrow {\rm B}).
  \end{array} \right.
\end{equation}
Consequently, we get
$\mathcal{S}_{\rm mix} ({\bf r}_0^N, \beta) \simeq - \ln \left( \frac{1}{N!} N_{\rm A}! N_{\rm B}!  \right) =- N ( X_{\rm A} \ln X_{\rm A} + X_{\rm B} \ln X_{\rm B} )= S_{\rm mix}^{(M=2)}$.
We numerically confirm this argument for binary hard sphere mixtures below.

The above argument can easily be generalised to a finite $M$-components systems.

\subsection{Numerical test}

We test the above argument numerically for $N=1000$ binary hard spheres in three dimensions by changing the concentration of the species A, $X_{\rm A}$, and the size ratio $R=\sigma_{\rm B}/\sigma_{\rm A}$.
We measure $\mathcal{S}_{\rm mix}$ by the method explained in Sec.~\ref{sec:numerical}.

\begin{figure}
\includegraphics[width=0.48\columnwidth]{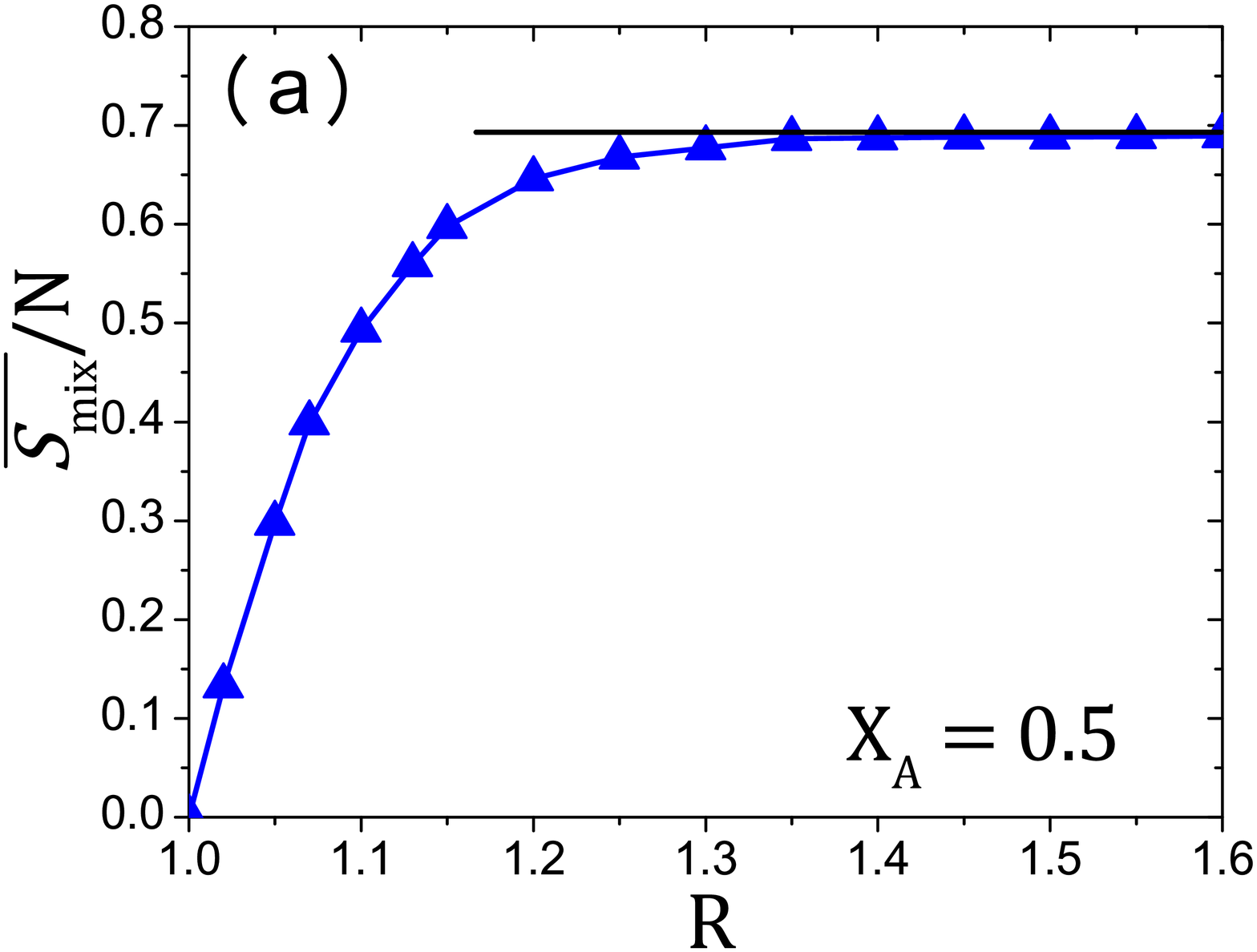}
\includegraphics[width=0.48\columnwidth]{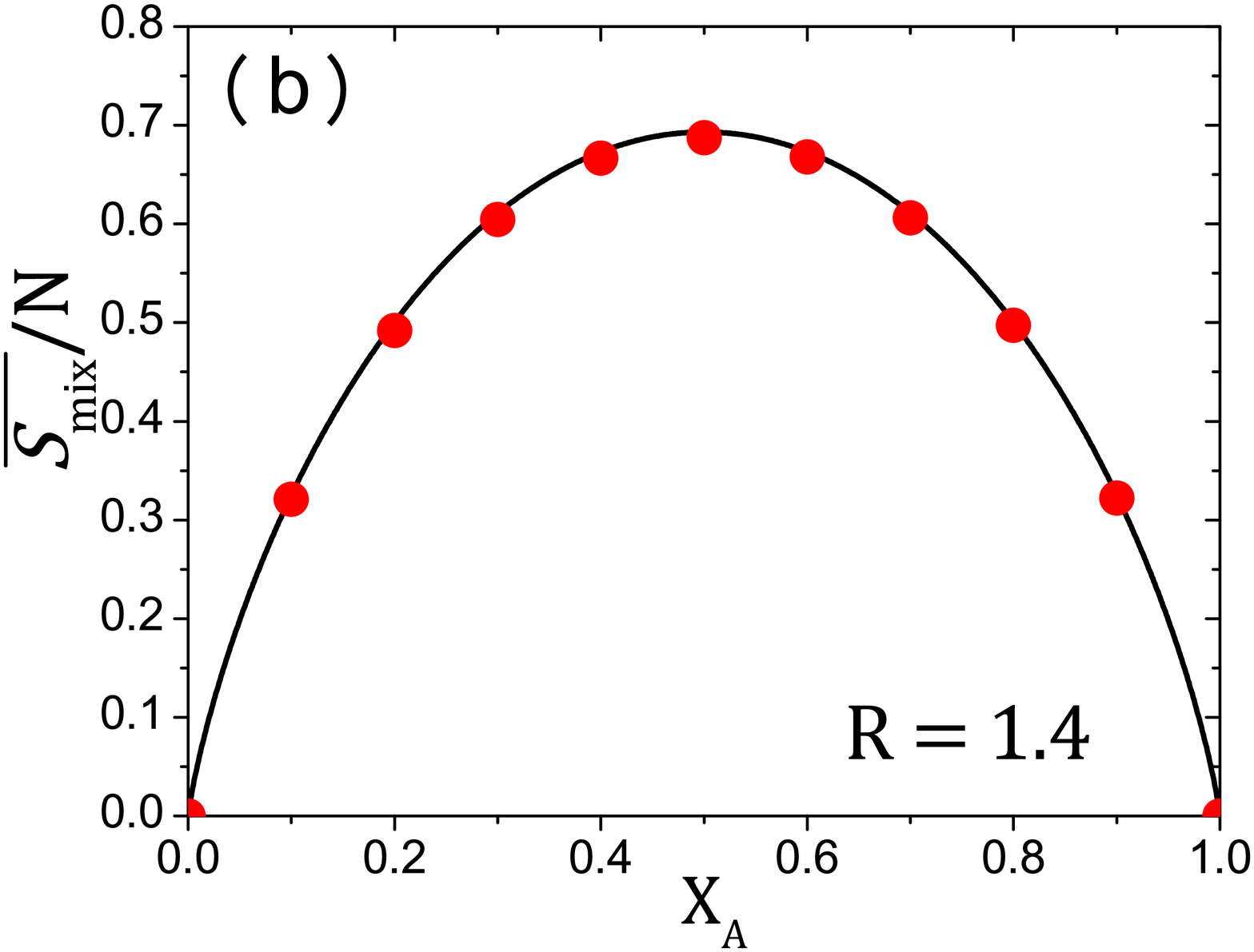}
\caption{Mixing entropy 
$\overline{\mathcal{S}_{\rm mix}}$ for three-dimensional binary hard sphere mixtures at $\phi=0.45$. (a) Evolution as a function of $R=\sigma_{\rm B}/\sigma_{\rm A}$ for $X_{\rm A}=X_{\rm B}=0.5$. The black straight line corresponds to $S_{\rm mix}^{(M=2)}/N =- X_{\rm A} \ln X_{\rm A} - X_{\rm B} \ln X_{\rm B} = \ln 2$. (b) Evolution as a function of $X_{\rm A}$ for $R=1.4$. The black curve corresponds to $S_{\rm mix}^{(M=2)}/N =- X_{\rm A} \ln X_{\rm A} - (1-X_{\rm A}) \ln (1-X_{\rm A})$.}
\label{fig:s_mix_binary}
\end{figure}

Figure~\ref{fig:s_mix_binary}(a) shows $\overline{\mathcal{S}_{\rm mix}}/N$ for equimolar mixtures ($X_{\rm A}=X_{\rm B}=0.5$) at $\phi=0.45$ as a function of $R$. 
As expected, $\overline{\mathcal{S}_{\rm mix}}/N$ vanishes in the monodisperse limit, $R \to 1$.
On the other hand, for $R \gtrsim 1.3$, $\overline{\mathcal{S}_{\rm mix}}/N$ converges to $S_{\rm mix}^{(M=2)}/N =- X_{\rm A} \ln X_{\rm A} - X_{\rm B} \ln X_{\rm B} = \ln 2$ indicated by the horizontal straight line.
Thus, we numerically confirm $\overline{\mathcal{S}_{\rm mix}}=S_{\rm mix}^{(M=2)}$ for binary mixtures with sufficiently large size ratio, and the monodisperse limit discussed in the above.
Furthermore, our numerical measurement smoothly connects the two cases around $1 \lesssim R \lesssim 1.3$.
Thus we no longer need to take any arbitrary decision about the mixing entropy~\cite{coluzzi1999thermodynamics} of any given physical system.

We find the above trend ($\overline{\mathcal{S}_{\rm mix}}/N \to 0$ at $R \to 1$ and $\overline{\mathcal{S}_{\rm mix}}=S_{\rm mix}^{(M=2)}$ for larger $R$) for larger volume fraction, $\phi \gtrsim 0.45$.
Since the $R \simeq 1$ region is difficult to study for $\phi \gtrsim 0.5$ due to crystallization, we show the data at $\phi=0.45$. It is likely that the crossover between monodisperse and bidisperse limits occurs at a smaller $R$ value when $\phi$ increases. 

We also measure the $X_{\rm A}$-dependence of $\overline{\mathcal{S}_{\rm mix}}$ for $R=1.4$ in Fig.~\ref{fig:s_mix_binary}(b).
We thus confirm that $\overline{\mathcal{S}_{\rm mix}}$ precisely follows the expected expression, $S_{\rm mix}^{(M=2)}/N =- X_{\rm A} \ln X_{\rm A} - (1-X_{\rm A}) \ln (1-X_{\rm A})$, when changing $X_{\rm A}$ systematically.

\section{$\mathcal{S}_{\rm mix}$ for hard sphere potential}
\label{sec:s_mix_hard}

For hard sphere potentials, the potential energy for the thermodynamic integration in Eq.~(\ref{eq:s_mix_integral}) is not a suitable observable.
Thus, we use the following numerical technique for this specific case. Conventionally, hard sphere systems are described by using the following pair potential $v_{i j}$ between particle $i$ and $j$, 
\begin{eqnarray}
v_{i j}(r_{i j}) &=& \left \{ \begin{array}{ll}
    \infty & (r_{i j} \leq \sigma_{i j} ), \\
    0 & (r_{i j} > \sigma_{i j} ),
  \end{array} \right. \nonumber \\
\beta &=& 1,
\end{eqnarray}
where $r_{i j}=| {\bf r}_i - {\bf r}_j |$, $\sigma_{i j}=(\sigma_i + \sigma_j)/2$.

Equivalently, we can adopt the following modeling by using a finite potential $\tilde{v}_{i j}$ but fixing instead $\beta=\infty$: 
\begin{eqnarray}
\tilde{v}_{i j}(r_{i j}) &=& \left \{ \begin{array}{ll}
    1 & (r_{i j} \leq \sigma_{i j} ), \\
    0 & (r_{i j} > \sigma_{i j} ),
  \end{array} \right. \nonumber \\
\beta &=& \infty. 
\label{eq:hard_sphere_potential_new} 
\end{eqnarray}

Thus, we perform the thermodynamic integration of Eq.~(\ref{eq:s_mix_integral}) using $\Delta U_{\rm mix}$ from $\beta=0$ to $\beta=\infty$ for the hard sphere systems described by Eq.~(\ref{eq:hard_sphere_potential_new}).

\bibliography{s_conf_poly}

\end{document}